\documentclass[prd,twocolumn,showpacs,amsmath,amssymb,floatfix]{revtex4}
\usepackage{graphicx,color,dcolumn,booktabs}
\usepackage{longtable,lscape}
\usepackage{txfonts}
\usepackage{amssymb}
\usepackage{indentfirst}
\usepackage{epsfig}

\def\beq{\begin{equation}}
\def\enq{\end{equation}}
\def\beqa{\begin{eqnarray}}
\def\enqa{\end{eqnarray}}

\def\nn{\nonumber}

\def\GeV{\nobreak\,\mbox{GeV}}
\newcommand{\rag}{\rangle}
\newcommand{\lag}{\langle}
\def\lb{\label}
\def\mmo{m_{D_sD_{s0}}}
\def\qq{\lag\bar{q}q\rag}
\def\sss{\lag\bar{s}s\rag}
\def\qqs{\lag\bar{s}s\rag}

\def\Gd{\lag g^2G^2\rag}
\def\G3{\lag g^3G^3\rag}
\def\alma{\alpha_{max}}
\def\almi{\alpha_{min}}
\def\bemi{\beta_{min}}
\def\al{\alpha}
\def\be{\beta}
\def\si{\sigma}

\begin{document}
\title{QCD sum rule calculation for the charmonium-like structures in 
the $J/\psi\phi$ and $J/\psi\omega$ invariant mass spectra}

\author{Stefano I. Finazzo}
\email{stefanofinazzo@gmail.com}
\affiliation{Instituto de F\'{\i}sica, Universidade de S\~{a}o Paulo,
C.P. 66318, 05315-970 S\~{a}o Paulo, SP, Brazil}
\author{Xiang Liu$^{1,2}$} 
\email{xiangliu@lzu.edu.cn}
\affiliation{$^1$School of Physical Science and Technology, Lanzhou 
University, Lanzhou 730000,  China\\
$^2$Research Center for Hadron and CSR Physics, Lanzhou University
$\&$ Institute of Modern Physics of CAS, Lanzhou 730000, China}
\author{Marina Nielsen}
\email{mnielsen@if.usp.br}
\affiliation{Instituto de F\'{\i}sica, Universidade de S\~{a}o Paulo,
C.P. 66318, 05315-970 S\~{a}o Paulo, SP, Brazil}

\date{\today}
\begin{abstract}

Using the QCD  sum rules we test if the charmonium-like structure 
$Y(4274)$, observed in the $J/\psi\phi$ invariant mass spectrum, can 
be described with a $D_s\bar{D}_{s0}(2317)+h.c.$ molecular current 
with $J^{PC}=0^{-+}$. We consider the contributions of condensates 
up to dimension ten and we work at leading order in $\alpha_s$.
We keep terms which are linear in the strange quark mass $m_s$. 
The mass obtained for such state is $m_{D_s{D}_{s0}}=(4.78\pm 0.54)$ 
GeV. We also consider a molecular $0^{-+}$ $D\bar{D}_{0}(2400)+h.c.$ 
current and we obtain $m_{D{D}_0}=(4.55\pm 0.49)$ GeV.
Our study shows that the newly observed $Y(4274)$ in the $J/\psi\phi$ 
invariant mass spectrum can be, considering the 
uncertainties, described using a molecular charmonium current.

\end{abstract}

 \maketitle

In the recent years, many new charmonium states were observed  by BaBar, 
Belle and CDF Collaborations. There is growing evidence that at least some 
of these new  states are non conventional $c\bar{c}$ states. In
some cases the masses of these states are very close to the 
meson-meson threshold, like the $X(3872)$ \cite{belle1}
and the $Z^+(4430)$ \cite{belle2}. Therefore, a molecular interpretation
for these states seems natural. Other possible interpretations for these 
states are  tetraquarks, hybrid mesons, or threshold effects. 

Very recently the CDF Collaboration \cite{Aaltonen:2011at} reported
a further study of the structures in the $J/\psi \phi$ invariant mass,
produced in exclusive $B^+\to J/\psi\phi K^+$ decays. Besides
confirming the  $Y(4140)$ state \cite{Aaltonen:2009tz} with a significance 
greater than 5$\sigma$, CDF also find evidence for a second structure 
with approximately $3.1\sigma$ significance. The reported mass and
width of this structure are $M=4274.4^{+8.4}_{-6.7}(\mathrm{stat})$ MeV 
and $\Gamma=32.3^{+21.9}_{-15.3}(\mathrm{stat})$ MeV
\cite{Aaltonen:2011at}. This new structure, refered as
$Y(4274)$ in ref.~\cite{Liu:2010hf}, was interpreted as the
S-wave $D_s\bar{D}_{s0}(2317)+h.c.$ molecular state. The authors of 
ref.~\cite{Liu:2010hf} have also predicted a S-wave $D\bar{D}_{0}(2400)+h.c.$
molecular state with a mass around 4.2 GeV,  which they call as the 
cousin of $Y(4274)$. This state is compatible with the enhancement 
structure around 4.2 GeV observed in the $J/\psi\omega$ invariant mass 
spectrum from $B$ decay \cite{Abe:2004zs}.

These two pseudoscalar molecular states could be the analogue of the
$Y(4140)$ and $Y(3930)$ (reported by the Belle Collaboration 
\cite{Abe:2004zs} and confirmed by the BaBar Collaboration 
\cite{Aubert:2007vj}), that were interpreted, in ref.~\cite{Liu:2009ei}, as
 $D_s^*\bar{D}_s^*$ and $D^*\bar{D}^*$ scalar molecular states
respectively. Some interpretations for the $Y(4140)$ can be found in refs.~
\cite{Liu:2009ei,Liu:2008tn,Mahajan:2009pj,Wang:2009ue,Branz:2009yt,
Albuquerque:2009ak,Liu:2009iw,Ding:2009vd,Zhang:2009st,vanBeveren:2009dc,
Stancu:2009ka,Liu:2009pu,Wang:2009ry,Drenska:2009cd,Molina:2009ct}.

Here we use the QCD sum rules (QCDSR) \cite{svz,rry,SNB,Nielsen:2009uh}, to 
check the suggestion made by the authors of ref.~\cite{Liu:2009ei}. 
Therefore, we study the two-point function based on a  $D_s\bar{D}_{s0}$ 
 molecular current with $J^{PC}=0^{-+}$, to see if the 
new observed structure, the $Y(4274)$, can be interpreted as such 
molecular state. We also investigate the $D\bar{D}_{0}$ molecular current.
Previous calculations for the new charmonium states interpreted as
molecular or tetraquark states can be found at 
\cite{Albuquerque:2009ak,x3872,molecule,lee,bracco,rapha,z12,zwid,mix,
x4350,Narison:2010pd}.

A possible $D_s\bar{D}_{s0}$ molecular current with $J^{PC}=0^{-+}$ is
given by%
\beq
j={i\over\sqrt2}\left[(\bar{s}_a\gamma_5 c_a)(\bar{c}_b s_b)+(\bar{c}_a
\gamma_5 s_a)(\bar{s}_b c_b)\right]
\;,
\label{field}
\enq
where $a$ and $b$ are color indices. 

The  QCDSR approach is based on the two-point correlation function
\beq
\Pi(q)=i\int d^4x ~e^{iq.x}\lag 0
|T[j(x)j^\dagger(0)]|0\rag.
\lb{2po}
\enq
The sum rule is obtained by evaluating the correlation function in 
Eq.~(\ref{2po}) in two ways: in the OPE side and in the phenomenological 
side. In the OPE side we  work at leading order 
in $\alpha_s$ in the operators, we consider the contributions from 
condensates up to dimension ten and  we keep terms which are linear in
the strange quark mass $m_s$. In the  phenomenological side,
the correlation function is calculated by inserting intermediate states 
for the $D_s\bar{D}_{s0}$ molecular state. The coupling of the molecular 
state, $M$, to the current, $j$, in Eq.~(\ref{field}) can be parametrized in 
terms of the parameter $\lambda$
\beq
\lag 0 |
j|M\rag =\lambda.
\label{lam}
\enq

Although  there is no one to one correspondence between
the current and the state, since the current
in Eq.~(\ref{field}) can be rewritten in terms of a sum over tetraquark type
currents, by the use of the Fierz transformation, the 
parameter $\lambda$, appearing in Eq.~(\ref{lam}), gives a measure of the 
strength of the coupling between the current and the state. 
Besides, as shown in ref.~\cite{Nielsen:2009uh}, in the Fierz transformation of a
molecular current, each tetraquark component contributes with suppression 
factors that originate from picking up the correct Dirac and color indices.
This means that if the physical state is a molecular state, it would be 
best to choose a molecular type of current 
so that it has a large overlap with the physical state. 
Therefore, if the sum rule gives a mass and width consistent with the 
physical values, we can infer that the physical state has a structure well 
represented by the chosen current. 

Using Eq.~(\ref{lam}), the phenomenological side
of Eq.~(\ref{2po}) can be written as 
\beq
\Pi^{phen}(q^2)={\lambda^2\over
\mmo^2-q^2}+\int_{0}^\infty ds\, {\rho^{cont}(s)\over s-q^2}, \lb{phe} 
\enq
where the second term in the RHS of Eq.(\ref{phe}) denotes the contribution
of the continuum of the states with the same quantum numbers as the current.
As usual in the QCDSR method, it is
assumed that the continuum contribution to the spectral density,
$\rho^{cont}(s)$ in Eq.~(\ref{phe}), vanishes below a certain continuum
threshold $s_0$. Above this threshold, it is given by
the result obtained in the OPE side. Therefore, one uses the ansatz 
\cite{io1}
\beq
\rho^{cont}(s)=\rho^{OPE}(s)\Theta(s-s_0)\;.
\enq

In the OPE side the correlation function can be  written as a
dispersion relation:
\beq
\Pi^{OPE}(q^2)=\int_{4m_c^2}^\infty ds {\rho^{OPE}(s)\over s-q^2}\;,
\lb{ope}
\enq
where $\rho^{OPE}(s)$ is given by the imaginary part of the
correlation function: $\pi \rho^{OPE}(s)=\mbox{Im}[\Pi^{OPE}(s)]$.
After transferring the continuum contribution to 
the OPE side, and after performing a Borel transform,
 the sum rule for the state described by a  
$D_sD_{s0}$ pseudoscalar molecular current can be written as:
\beq \lambda^2e^{-\mmo^2/M^2}=\int_{4m_c^2}^{s_0}ds~
e^{-s/M^2}~\rho^{OPE}(s)\;, \lb{sr}
\enq
where
\beq
\rho^{OPE}(s)=\sum_{D=0}^{10}\rho^{[D]}(s)
\lb{rhoeq}
\enq
with $\rho^{[D]}$ representing the dimension-$D$ condensates. To extract 
the mass of the state we take the derivative of Eq.~(\ref{sr})
with respect to $1/M^2$, and divide the result by Eq.~(\ref{sr}):
\beq
\mmo^2={\int_{4m_c^2}^{s_0}ds~e^{-s/M^2}~s~\rho^{OPE}(s)\over
\int_{4m_c^2}^{s_0}ds~e^{-s/M^2}~\rho^{OPE}(s)}.
\enq

The contributions to $\rho^{OPE}(s)$, up to dimension-ten condensates, 
using factorization hypothesis, are given by:
\beqa\label{eq:pert}
&&\rho^{[0]}(s)={3\over 2^{11} \pi^6}\int\limits_{\almi}^{\alma}
{d\al\over\alpha^3}
\int\limits_{\bemi}^{1-\al}{d\be\over\be^3}(1-\al-\be)
\nn\\
&&\times\left[(\al+\be)m_c^2-\al\be s\right]^4,
\nn
\enqa
\beqa\label{eq:cond3}
&&\rho^{[3]}(s)={-3m_s\sss\over 2^{7}\pi^4}\int\limits_{\almi}^{\alma}
{d\al\over\al}\left\{-{(m_c^2-\al(1-\al)s)^2\over1-\al}\right.+
\nn\\
&&\left.4m_c^2\int\limits_{\bemi}^{1-\al}{d\be\over\be}\left[(\al+\be)m_c^2
-\al\be s\right]\right\},
\nn
\enqa
\beqa\label{eq:con4}
&&\rho^{[4]}(s)={\Gd\over2^{10}\pi^6}\int\limits_{\almi}^{
\alma}{d\al\over\alpha}\int\limits_{\bemi}^{1-\al}{d\be}\left[(\al+\be)m_c^2
-\al\be s\right]
\nn\\
&&\times\left\{{m_c^2(1-\al-\be)\over\al^2}+{3\over2}
{(\al+\be)m_c^2-\al\be s\over\be^2}\right\},
\nn
\enqa
\beqa\label{eq:cond5}
&&\rho^{[5]}(s)=-{m_sm_0^2\sss\over 2^{7}\pi^4}\left\{
(2m_c^2-s)\sqrt{1-4m_c^2/s}\right.
\nn\\
&&+3m_c^2\ln\left[1+\sqrt{1-4m_c^2/s}\over1-\sqrt{1-4m_c^2/s}\right]
\nn\\
&&\left.-6m_c^2\int\limits_{\almi}^{\alma}{d\al}\ln\left[ (1-\al)
\left({s\over m_c^2}-{1\over\al}\right)\right]\right\},
\nn
\enqa
\beqa\label{eq:cond6}
&&\rho^{[6]}(s)=-{m_c^2\sss^2\over 16\pi^2}\sqrt{1-4m_c^2/s}
\nn\\
&&+{\G3\over2^{12}\pi^6}\int\limits_{\almi}^{\alma}{d\al\over\alpha^3}\int\limits_
{\bemi}^{1-\al}{d\be}(1-\al-\be)\left[(\al+3\be)m_c^2
-\al\be s\right]
\nn
\enqa
\beqa\label{eq:cond8}
&&\rho^{[8]}(s)={m_c^2m_0^2\sss^2\over 2^5\pi^2}\int_0^1
d\al~\delta\left(s-{m_c^2\over \al(1-\al)}\right)
\nn\\
&&\times\left[{m_c^2\over \al(1-\al)M^2}-{1+\al\over1-\al}\right]+{\Gd^2
\over2^{14}\pi^6}\bigg[\sqrt{1-4m_c^2/s}
\nn\\
&&+{m_c^2\over9}\int_0^1 {d\al d\be\over\al^2}\Theta[1-(\al+\be)]\delta\left(s-
{(\al+\be)m_c^2\over \al\be}\right)
\nn\\
&&\times\left[-6+{(1-\al-\be)\over\be^2}{m_c^2\over M^2}
\right]\Bigg]
\nn
\enqa
\beqa\label{eq:cond10}
&&\rho^{[10]}(s)=-{(m_0^2\sss)^2\over 2^8\pi^2}\int_0^1
{d\al\over\al(1-\al)}~\delta\left(s-{m_c^2\over \al(1-\al)}\right)
\nn\\
&&\times\left[{m_c^6\over \al(1-\al)M^6}-{4m_c^4\over(1-\al)M^4}
+4{m_c^2\over M^2}\right]
\nn\\
&&+{\sss^2\Gd\over2^6\pi^2}\int_0^1
{d\al\over\al}~\delta\left(s-{m_c^2\over \al(1-\al)}\right){m_c^2\over M^2}
\bigg[-{1\over3\al}
\nn\\
&&+{m_c^2\over M^2}\left({1\over9\al^2}+{1\over(1-\al)^2}\right)-{m_c^4\over 
18M^4}{1\over\al(1-\al)^2}\bigg]
\nn\\
&&-{\Gd\G3\over3^22^{15}\pi^6}\int_0^1 {d\al d\be\over\al^2}\Theta[1-(\al+\be)]
\delta\left(s-{(\al+\be)m_c^2\over \al\be}\right)
\nn\\
&&\times\left[(1-\al-\be){m_c^2\over\be^2M^2}\left(2{m_c^2\over\al M^2}-1\right)
+3-6{m_c^2\over\al M^2}\right],
\enqa
where the integration limits are given by $\almi=({1-\sqrt{1-
4m_c^2/s})/2}$, $\alma=({1+\sqrt{1-4m_c^2/s})/2}$, $\bemi={\al
m_c^2/( s\al-m_c^2)}$, and we have used $\lag\bar{s}g\si.Gs\rag=m_0^2\sss$. 
For consistency, we have included the small contribution of the dimension-six
condensate $\langle g^3 G^3\rangle$.  We have also included 
the dimension-8 and  dimension-10 condensate contributions, 
related with the mixed condensate-quark condensate, gluon condensate squared, 
mixed condensate squared, four-quark condensate-gluon condensate and,
three-gluon condensate-gluon condensate.

For a consistent comparison with the results obtained for the other molecular
states using the QCDSR approach, we  have considered here the same values 
used for the quark masses and condensates  as in 
refs.~\cite{x3872,molecule,lee,bracco,rapha,z12,zwid,mix,
x4350,Narison:2010pd,narpdg}:
$m_c(m_c)=(1.23\pm 0.05)\,\GeV $, $m_s=(0.13\pm 0.03)\,\GeV $,
$\lag\bar{q}q\rag=\,-(0.23\pm0.03)^3\,\GeV^3$, $\qqs=0.8\qq$,
$m_0^2=0.8\,\GeV^2$, $\lag g^2G^2\rag=0.88~\GeV^4$. For the three-gluon
condensate we use $\lag g^3G^3\rag=0.045~\GeV^6$ \cite{svz}.

\begin{figure}[h]
\centerline{\epsfig{figure=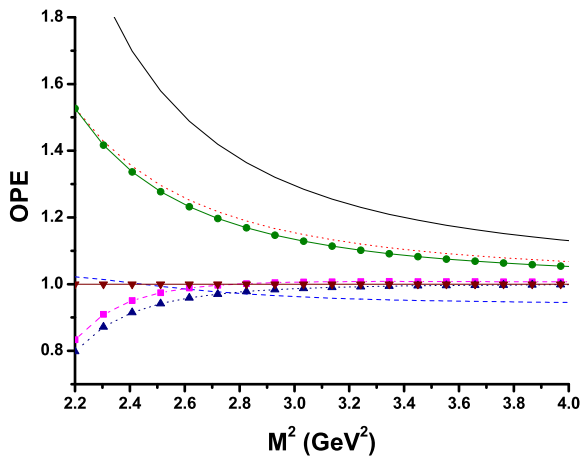,height=70mm}}
\caption{The OPE convergence for the $J^{PC}=0^{-+},~D_sD_{s0}$ 
molecule, in the region
$2.2 \leq M^2 \leq4.0~\GeV^2$ for $\sqrt{s_0} = 5.2$ GeV.  We plot the 
relative contributions starting with the perturbative contribution 
(solid line), and each other line represents the 
relative contribution after adding of one extra condensate in the expansion: 
+ $D=3$  (dashed line), 
+ $D=4$ (dotted line), + $D=5$ 
(solid line with circles), + $D=6$ (dashed line with squares), + $D=8$ 
(dotted line with triangles), + $D=10$ (solid line with triangles).}
\label{figconv}
\end{figure}

The continuum threshold is a physical parameter that should be determined from
the spectrum of the mesons. The value of the continuum threshold in the QCDSR 
approach is, in general, given as the value of the mass of the first excitated 
state squared. In some known cases, like the $\rho$ and $J/\psi$, 
the first excitated state has a mass approximately $0.5~\GeV$ above the
ground state mass.  In the cases that one does not know the spectrum, one 
expects the continuum threshold to be approximately the square of the mass 
of the state plus $0.5~\GeV$: $s_0=(m_0+0.5~\GeV)^2$. Therefore, to fix the 
continuum threshold range we extract the mass from the sum rule, for a given 
$s_0$, and accept such value of $s_0$ if the obtained mass is in the range 
0.4 GeV to 0.6 GeV smaller than $\sqrt{s_0}$. Using this criterion,
we obtain $s_0$ in the range $5.1\leq\sqrt{s_0}\leq 5.3~\GeV$.

The Borel window is determined by analysing the OPE convergence, the Borel
stability  and the pole contribution. To determine the minimum value of the 
Borel mass we impose that the contribution of the higher dimension 
condensate should be smaller than 10\% of the total contribution: 
$M^2_{min}$ is such that
\beq
\left|{\mbox{OPE summed up dim n-1 }(M^2_{min})\over\mbox{total 
contribution }(M^2_{min})}\right|=0.9.
\label{mmin}
\enq

\begin{figure}[h]
\centerline{\epsfig{figure=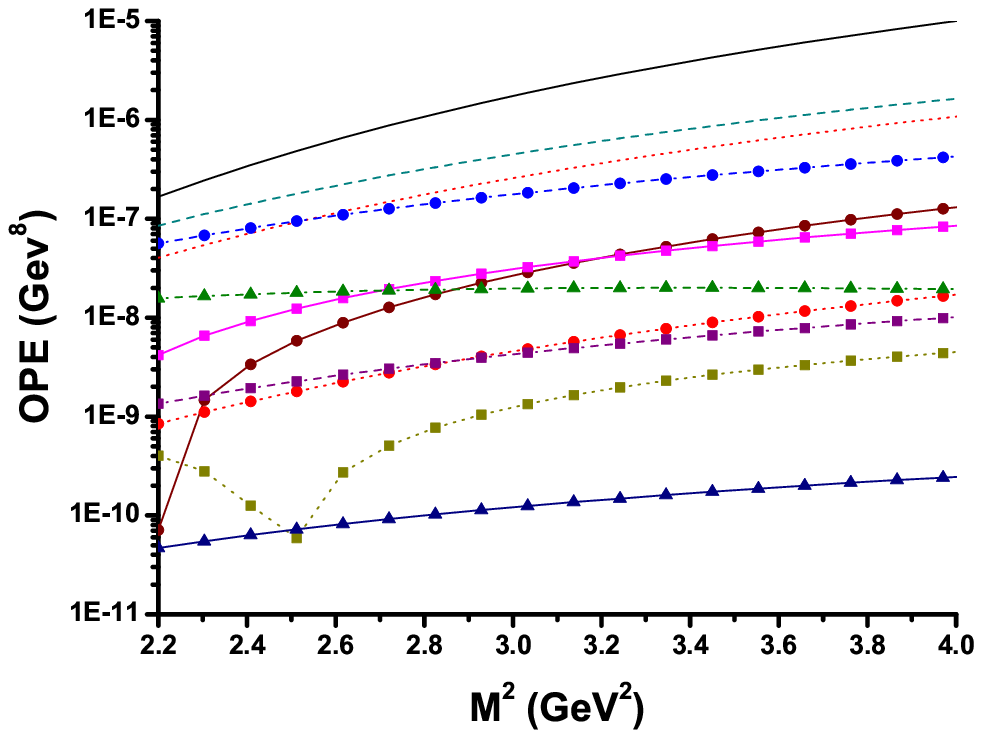,height=70mm}}
\caption{The OPE convergence for the $J^{PC}=0^{-+},~D_sD_{s0}$ 
molecule, in the region
$2.2 \leq M^2 \leq4.0~\GeV^2$ for $\sqrt{s_0} = 5.2$ GeV.  We plot the 
contributions of all individual condensates in the OPE: the perturbative 
contribution (solid line), $\sss$ contribution (dashed line), $\Gd$ contribution 
(dotted line), $m_0^2\sss$ (solid line with cicles), $\sss^2$ (dashed line 
with circles), $\G3$ (dotted line with circles), $m_0^2\sss^2$ (solid line with 
squares), $\Gd^2$ (dashed line with squares), $(m_0^2\sss)^2$ (dotted line with 
squares), $\sss^2\Gd$ (dashed line with triangles), $\Gd\G3$ (solid line with 
triangles).}
\label{opeterms}
\end{figure}

In Fig.~\ref{figconv} we show the relative contribution of all the terms 
in the
OPE side of the sum rule, in the region $2.2 \leq M^2 \leq4.0~\GeV^2$, for 
$\sqrt{s_0} = 5.2~\GeV$. From this figure we see  that the 
contribution of the dimension-10 condensate is smaller than 10\% of the total 
contribution for values of $M^2\geq2.4~\GeV^2$, and that we have an excellent 
OPE convergence for $M^2\geq2.4~\GeV^2$. To have an idea of the importance
of the different terms in the OPE, we show, in Fig.~\ref{opeterms}, the
contribution of each condensate. As we can see, the condensates of dimension
higher than six are, at least, one order of magnetude smaller than the 
perturbative contribution, in all considered Borel region.

\begin{figure}[h]
\centerline{\epsfig{figure=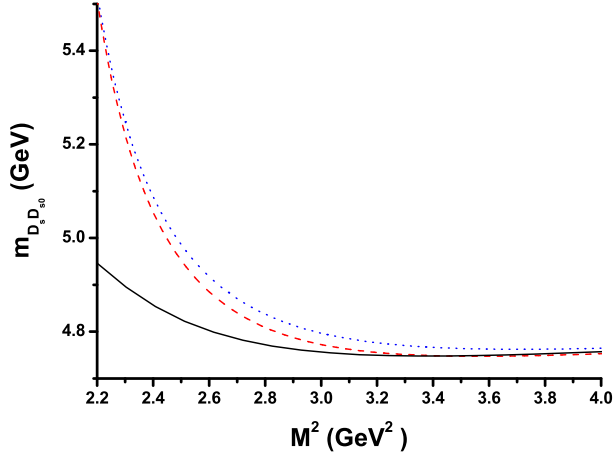,height=70mm}}
\caption{The pseudoscalar meson mass, described with a $D_sD_{s0}$ molecular 
current, as a function of the sum rule parameter ($M^2$) for $\sqrt{s_0} =
5.2$ GeV. The solid line shows the result obtained considering all
contributions up to dimension-10. The dashed and dotted lines show, respectively,
 the results obtained neglecting the contributions of the dimension-8 
($\Gd^2$) and dimension-10 ($\Gd\G3,~\Gd\sss^2$) gluon condensates.}
\label{figm1}
\end{figure}

As commented above, the OPE convergence is very good in the Borel range
$2.4 \leq M^2 \leq4.0~\GeV^2$. However, the Borel stability
for the mass of the state is only good for $M^2\geq2.8~\GeV^2$, as can be 
seen through the solid line in Fig.~\ref{figm1}. Therefore, we  fix the lower
value of $M^2$ in the sum rule window as $M^2_{min}= 2.8$ GeV$^2$.

\begin{figure}[h]
\centerline{\epsfig{figure=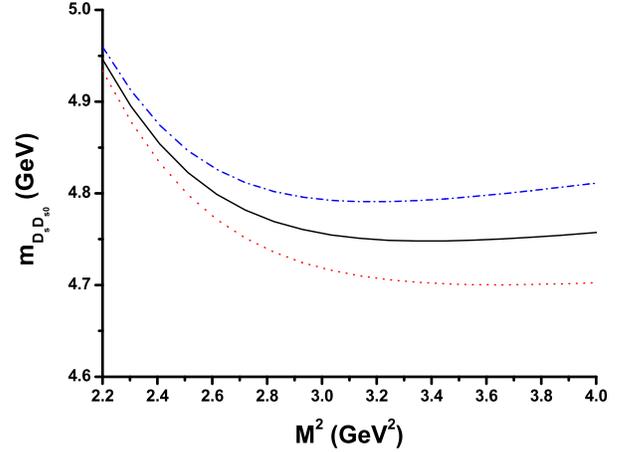,height=70mm}}
\caption{The pseudoscalar meson mass, described with a $D_sD_{s0}$ molecular 
current, as a function of the sum rule parameter
($M^2$) for $\sqrt{s_0} =5.1$ GeV (dotted line), $\sqrt{s_0} =5.2$ GeV 
(solid line) and $\sqrt{s_0} =5.3$ GeV (dot-dashed line).}
\label{figm2}
\end{figure}

In Fig.~\ref{figm1} we also show, through the dashed and dotted lines, the result 
obtained if we neglect the contribution of dimension-8 and dimension-10
gluon condensates. We see that the contribution of the dimension-8 and -10 gluon
condensates ($\Gd^2,~\Gd\G3,~\Gd\sss^2$) are only important in the
region $M^2\leq2.8\GeV^2$, which is not in our Borel window, due
to the mass stability. Therefore, at least in this case, the contribution
of higher dimension gluon condensates could be safely neglected. Besides,
as can be seen in  Fig.~\ref{figm2}, where we show the results for the mass
for different values of $\sqrt{s_0}$ considering all condensate contributions 
up to dimension-10, the dependence of the mass on the OPE 
convergence is smaller than the its dependence on the continuum threshold 
parameter.

To be able to extract, from the sum rule, information about the low-lying 
resonance, the pole contribution to the sum rule should be bigger than, 
or at least equal to, the continuum contribution. Since the continuum 
contribution increases with $M^2$, due to the dominance of the perturbative 
contribution, we fix the maximum value of the Borel mass
to be the one for which the pole contribution is equal to the continuum 
contribution. 

\begin{figure}[h]
\centerline{\epsfig{figure=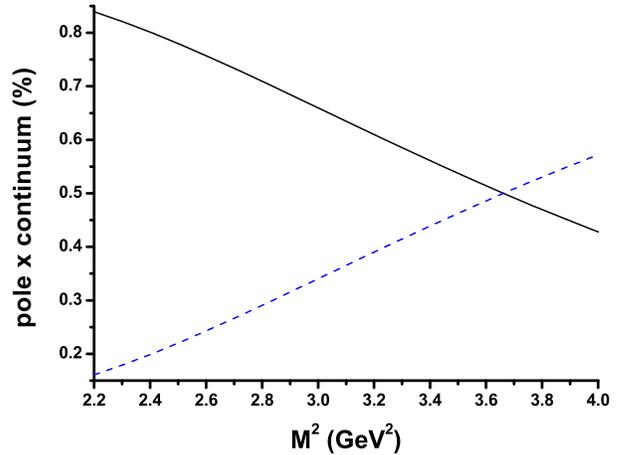,height=70mm}}
\caption{The solid line shows the relative pole contribution (the
pole contribution divided by the total, pole plus continuum,
contribution) and the dashed line shows the relative continuum
contribution for $\sqrt{s_0}=5.2~\GeV$.}
\label{figpvc}
\end{figure}

From Fig.~\ref{figpvc} we see that for $\sqrt{s_0}=5.2~\GeV$, the pole 
contribution is bigger than the continuum contribution for $M^2\leq3.66
\GeV^2$. We show in Table I the values of $M_{max}$ for other values 
of $\sqrt{s_0}$. Although for $\sqrt{s_0}=5.0~\GeV$ there is still a small
 allowed Borel window, the difference between the obtained mass and the
continuum threshold is very small (smaller than 0.2 GeV). Therefore,
we do not consider values of $\sqrt{s_0}<5.1~\GeV$.

\begin{center}
\small{{\bf Table I:} Upper limits in the Borel window for the $0^{-+},~
D_sD_{s0}$ current 
obtained from the sum rule for different values of $\sqrt{s_0}$.}
\\
\begin{tabular}{|c|c|}  \hline
$\sqrt{s_0}~(\GeV)$ & $M^2_{max}(\GeV^2)$  \\
\hline
 5.1 & 3.43 \\
\hline
 5.2 & 3.66 \\
\hline
5.3 & 3.90 \\
\hline
\end{tabular}\end{center}

To estimate the dependence of our results with the values of the quark 
masses and condensates, we fix $\sqrt{s_0}=5.2~\GeV$ and vary the other 
parameters in the ranges: $m_c=(1.23\pm0.05)~\GeV$, $m_s=(0.13\pm 0.03)\,
\GeV $, $\lag\bar{q}q\rag=\,-(0.23\pm0.03)^3\,\GeV^3$, 
$m_0^2=(0.8\pm0.1)\GeV^2$. 
In our calculation we have assumed the factorization hypothesis. However, it 
is important to check how a violation of the factorization hypothesis would
modify our results. To do that we multiply the contribution of the four-quark
condensates of $D=6,8$ and $D=10$ in Eq.~(\ref{eq:cond10}) 
by a factor $K$, and we vary $K$ in the range $0.5\leq K\leq2$. The 
dependence of our results with all the variations mentioned above is show in
Table II.

\begin{center}
\small{{\bf Table II:} Values obtained for $\mmo$, in the Borel window 
$3.0\leq M^2\leq 3.65~\GeV^2$, when the parameters vary in the ranges
showed.}
\\
\begin{tabular}{|c|c|}  \hline
parameter & $\mmo~(\GeV)$  \\
\hline
$m_c=(1.23\pm0.05)~\GeV$  & $4.76\pm0.07$ \\
\hline
$m_s=(0.13\pm 0.03)\,\GeV $ &$4.76\pm0.06$  \\
\hline
$\lag\bar{q}q\rag=\,-(0.23\pm0.03)^3\,\GeV^3$ & $4.89\pm0.27$ \\
\hline
$m_0^2=(0.8\pm0.1)\GeV^2$ & $4.753\pm0.003$ \\
\hline
$0.5\leq K\leq2$ & $4.80\pm0.11$ \\
\hline
\end{tabular}\end{center}

Taking into account the uncertainties given above and the uncertainties
due to the continuum threshold parameter and due to the OPE convergence, 
we finally arrive at
\beq
\mmo = (4.78\pm0.54)~\GeV,
\label{ymass}
\enq
which, considering the error, is still in agreement with the mass of the 
 newly observed structure $Y(4274)$.

One can also deduce, from Eq.~(\ref{sr}), the parameter 
$\lambda$ defined in Eq.~(\ref{lam}). We get:
\beq
\lambda = \left(6.0\pm3.9\right)\times 10^{-2}~\GeV^5.
\label{la1}
\enq
This number is of the same order as the current-state coupling obtained
in ref.~\cite{Albuquerque:2009ak}, where the $J^{PC}=0^{++}$ $D_s^*D_s^*$
molecular current was considered to describe the $Y(4140)$:
\beq
\lambda_{D_s^*D_s^*} = \left(4.22\pm0.83\right)\times 10^{-2}~\GeV^5.
\label{lads}
\enq
Therefore, we can conclude that the state can be well represented by the
$J^{PC}=0^{-+}$ $D_s\bar{D}_{s0}$ molecular current

\begin{figure}[h]
\centerline{\epsfig{figure=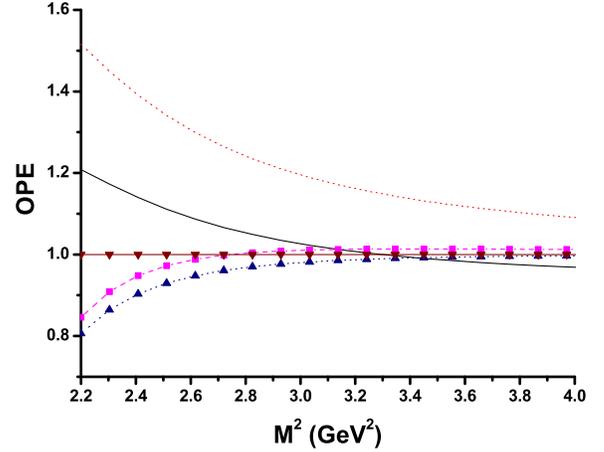,height=70mm}}
\caption{Same as Fig. 1 for the $DD_0$ current for $\sqrt{s_0}=5.0~\GeV$:
perturbative contribution (long-dashed line), 
relative contribution after adding of one extra condensate in the expansion: 
+ $D=4$ (dotted line), + $D=6$ (dashed line with squares), + $D=8$ 
(dotted line with triangles), + $D=10$ (solid line with triangles).}
\label{ope-dd0}
\end{figure}

\begin{figure}[h]
\centerline{\epsfig{figure=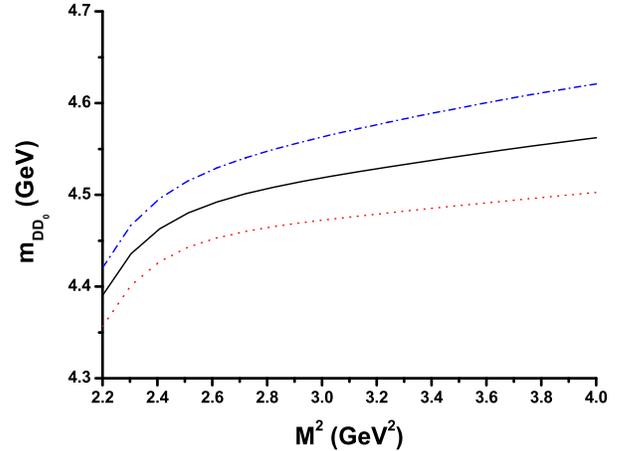,height=70mm}}
\caption{Same as Fig.~\ref{figm2} for the $DD_{0}$ molecular 
current, for $\sqrt{s_0} =4.9$ GeV (dotted line), $\sqrt{s_0} =5.0$ GeV 
(solid line) and $\sqrt{s_0} =5.1$ GeV (dot-dashed line).}
\label{mdd0}
\end{figure}

To obtain results for the $D\bar{D}_0$  molecular current with 
$J^{PC}=0^{-+}$, we only have to take $m_s=0$ and $\sss=\qq$ in 
Eq.~(\ref{eq:cond10}).  As can be seen by  
Fig.~\ref{ope-dd0}, the OPE convergence in this case is also very good
for $M^2\geq2.4~\GeV^2$. Therefore to fix the minimum value of the
Borel parameter, we will consider the Borel stability of the obtained mass.
For this we show, in Fig.~\ref{mdd0} the results for the mass of the
state described by a $DD_{0}$ pseudoscalar molecular current, for different
values of $\sqrt{s_0}$. We see that for $M^2\geq2.7~\GeV^2$ we get a good
Borel stability. Therefore we fix $M^2_{min}=2.7~\GeV^2$.

In Table III we give the values of $M_{max}$ for the considered values 
of $\sqrt{s_0}$. 

\begin{center}
\small{{\bf Table III:} Upper limits in the Borel window for the $0^{-+},~
DD_{0}$ current obtained from the sum rule for different values of 
$\sqrt{s_0}$.}
\\
\begin{tabular}{|c|c|}  \hline
$\sqrt{s_0}~(\GeV)$ & $M^2_{max}(\GeV^2)$  \\
\hline
 4.9 & 3.19 \\
\hline
 5.0 & 3.40 \\
\hline
5.1 & 3.61 \\
\hline
\end{tabular}\end{center}

Taking into account the uncertainties due to the quark masses, condensates,
continuum threshold parameter and OPE convergence, we finally arrive at
\beq
m_{DD_0} = (4.55\pm0.49)~\GeV,
\label{ymass}
\enq
which, although a little bigger than the prediction in 
ref.~\cite{Liu:2010hf} for a S-wave $D\bar{D}_{0}$ molecular state,
is still in agreement with it, considering the error. It is interesting to 
notive that the result in Eq.~(\ref{ymass}) is in a excellent agreement with
the result obtained in ref.~\cite{Chen:2010jd}, where different tetraquark
currents were used to study  $J^{PC}=0^{--}$ and $0^{-+}$ charmonium-like 
states. For the parameter $\lambda$ we get:
\beq
\lambda_{DD_0} = \left(5.4\pm3.9\right)\times 10^{-2}~\GeV^5.
\label{ladd0}
\enq
 
The mass we have obtained for the $D\bar{D}_{0}$ molecular state is
approximately two hundred MeV below than the value obtained for the similar 
strange state. This is very different from the results obtained in
ref.~\cite{Albuquerque:2009ak} where the $J^{PC}=0^{++}$ $D_s^*D_s^*$
and $D^*D^*$ molecular currents were considered. In the case of the scalar
molecular currents, the difference between the masses of the strange and 
non-strange states was consistent with zero.

In conclusion, the newly observed structure $Y(4274)$ in the
$J/\psi\phi$ invariant mass spectrum can be, considering the errors,
 interpreted as the S-wave $D_s\bar{D}_{s0}+h.c.$ molecular 
charmonium, in agreement with the findings in ref.~ \cite{Liu:2010hf}, where 
a dynamical study of the system, composed of the pseudoscalar and
scalar charmed mesons, was done. In the case of the  S-wave $D\bar{D}_{0}
+h.c.$ molecular current, which was called as the cousin of $Y(4274)$
in ref.~ \cite{Liu:2010hf}, the QCDSR results are 
consistent with the enhancement structure around 4.2 GeV in the
$J/\psi\omega$ invariant mass spectrum from $B$ decay
\cite{Abe:2004zs}.

\section*{Acknowledgment}

\noindent 
 This work has been partly supported by FAPESP and CNPq-Brazil, and by
the National Natural Science Foundation of
China under Grants No.
11035006, No. 11047606 and the Ministry of Education of China
(FANEDD under Grant No. 200924, DPFIHE under Grants No.
20090211120029, NCET under Grant No. NCET-10-0442, the Fundamental
Research Funds for the Central Universities.

\vfil

\end{document}